\documentclass[aps, twocolumn, superscriptaddress]{revtex4}
\usepackage{graphicx, epsfig}
\usepackage{epstopdf}

\DeclareGraphicsExtensions{.  jpg, .  pdf,  .  mps,  . png,  . eps,  .  ps,  .  EPS}
\usepackage{amsmath}
\usepackage{color}
\begin{document}
\newcommand{\avg}[1]{\langle{#1}\rangle}
\newcommand{\ket}[1]{\left |{#1}\right \rangle}

\newcommand{\half}{\mbox{$\textstyle \frac{1}{2}$}}
\newcommand{\proj}[1]{\ket{#1}\bra{#1}}
\newcommand{\av}[1]{\langle #1\rangle}
\newcommand{\braket}[2]{\langle #1 | #2\rangle}
\newcommand{\bra}[1]{\langle #1 | }
\newcommand{\Avg}[1]{\left\langle{#1}\right\rangle}
\newcommand{\inprod}[2]{\braket{#1}{#2}}
\newcommand{\upket}{\ket{\uparrow}}
\newcommand{\downket}{\ket{\downarrow}}
\newcommand{\Tr}{\mathrm{Tr}}
\newcommand{\hcontrol}{*!<0em, . 025em>-=-{\Diamond}}
\newcommand{\hctrl}[1]{\hcontrol \qwx[#1] \qw}
\newenvironment{proof}[1][Proof]{\noindent\textbf{#1. } }{\ \rule{0. 5em}{0. 5em}}
\newtheorem{mytheorem}{Theorem}
\newtheorem{mylemma}{Lemma}
\newtheorem{mycorollary}{Corollary}
\newtheorem{myproposition}{Proposition}
\newcommand{\vp}{\vec{p}}
\newcommand{\Or}{\mathcal{O}}
\newcommand{\so}[1]{{\ignore{#1}}}

\newcommand{\red}[1]{\textcolor{red}{#1}}
\newcommand{\blue}[1]{\textcolor{blue}{#1}}

\newcommand{\bea}{\begin{eqnarray}}
\newcommand{\eea}{\end{eqnarray}}
\newcommand{\gt}{\tilde{g}}
\newcommand{\mt}{\tilde{\mu}}
\newcommand{\et}{\tilde{\varepsilon}}
\newcommand{\ct}{\tilde{C}}
\newcommand{\bt}{\tilde{\beta}}

\title{Weighted Growing Simplicial Complexes}

\author{Owen Courtney}
\affiliation{School of Mathematical Sciences, Queen Mary University of London, E1 4NS, London, UK}
\author{Ginestra Bianconi}
\affiliation{School of Mathematical Sciences, Queen Mary University of London, E1 4NS, London, UK}

\begin{abstract}
Simplicial complexes describe collaboration networks, protein interaction networks and brain networks and in general network structures in which the interactions can include more than two nodes. In real applications, often simplicial complexes are weighted.
Here we  propose a non-equilibrium  model for  weighted growing simplicial complexes. The proposed dynamics is able to generate  weighted simplicial complexes with  a rich  interplay between weights and topology emerging not just at the level of nodes and links, but also at the level of faces of higher dimension.
\end{abstract}

\pacs{89.75.-k,89.75.Fb,89.75.Hc}

\maketitle
\section{Introduction}
Recently generalized network structures such as multilayer networks \cite{PhysReports,Alex_review} and simplicial complexes \cite{Kahle,Bassett,EPL} are attracting increasing attention in the network science community.  
In the last twenty years very significant advances in the understanding of complex systems have been  obtained using network theory \cite{NS,Doro_book,Newman_book}.
 Since the increasingly rich Big Data datasets on social, technological and biological systems often  include information that goes beyond what is possible to describe by a single network it is now becoming of fundamental importance to characterize generalized network structures. For instances in many cases data include a set of  interactions having different connotations  or   occurring between more than two nodes.
The flourishing field of multilayer networks \cite{PhysReports,Alex_review} defines a way to treat networks where links have different connotations.
Simplicial complexes \cite{Kahle,Bassett,EPL}, instead,  allow for the description of interactions occurring between more than two nodes. As such simplicial complexes are important to study a large variety of complex systems.
For example in scientific collaboration networks collaboration extend to teams of more than two scientists, or in actor collaboration networks, rarely a cast of a movie includes exclusively two actors.
In on-line social networks the rich structure of possible actions such as tagging, posting or linking to other users also allows for the identification of interactions between more than two nodes. In biology simplicial complexes are useful to undestand protein interaction networks. In fact, in order to perform a function proteins in the cell bind together to form protein complexes typically including more than two different types of interacting protein. Finally simplicial complexes are becoming increasingly popular in analysing brain networks \cite{Bassett} where one needs to distinguish for instance between three regions of the brain which are  pairwise correlated, and the case in which they are typically activated all at the same time. 

From the network perspective simplicial complexes can be  interchanged with hypergraphs \cite{Newman1,Zlatic} for the analysis of real networked datasets. However  simplicial complexes also have a geometrical interpretation  and they can be interpreted as the result of gluing nodes, links, triangles, tetrahedra etc. along their faces. As such  simplicial complexes can be used to characterize  the resulting network geometry   using for instance  novel definitions of network curvatures \cite{Ollivier,Gromov,Yau,Reka,Jost, Arejit1,Arejit2,Nott,Caldarelli} or characterizing their emergent geometrical properties \cite{Emergent, Hyperbolic}. Simplicial complexes are also starting to be widely used to perform a topological analysis of network datasets and of dynamical processes defined on networks  \cite{Bassett,Vaccarino1, Vaccarino2,Taylor}. Most notably this approach has been applied to brain functional networks showing that these novel  techniques can reveal important differences between networked datasets that cannot be detected by more traditional methods \cite{Bassett,Vaccarino1,Vaccarino2}.

Different frameworks have been proposed to model simplicial complexes.
On one side there are equilibrium models of static simplicial complexes that generalize the random graph or the configuration model to simplicial complexes \cite{Zuev, Farber1, Farber2, Owen1, Kahle, Newman1,Zlatic}.
On the other side there are non-equilibrium models of growing simplicial complexes displaying emergent structural properties and  geometry \cite{Emergent,CQNM,FB,NGF,Hyperbolic}. These models generalize at the same time growing network models \cite{BA,Doro_growing} with preferential attachment and Apollonian networks \cite{Ap1,Ap2,Ap3,Ap4}.

In real applications, simplicial complexes  are  typically weighted which explains the need to extend the modelling framework to characterize weighted simplicial complexes. For instance in scientific collaboration networks teams of collaborators can be weighted by the strength of their collaboration (how many papers a scientific collaboration has produced).
Here we characterize a weighted simplicial network model using a non-equilibrium dynamics. The evolution of the topology of these networks is based on the recently proposed framework of Network Geometry with Flavor \cite{NGF} able to generate networks with different complex topology, including hyperbolic manifolds, scale-free networks, and networks with relevant modularity.

Here our focus will be on characterizing the rich interplay between weights and topology in these models.
In single networks \cite{Alain,Weighted,Boguna} it has been shown by analysis of a vast set of real datasets that weights might not be distributed uniformly over the links of the network. 
Specifically in some networks, hubs nodes can have connections with on average stronger weights than the typical connections of low degree nodes. The way to characterize these weight-topology correlations is by studying the scaling of the average strength of nodes as a function of their degree.
If the strength grows linearly with the degree the weights are uniformly distributed among the nodes of the network. If, instead the observed scaling is superlinear then hubs typically have links with stronger weights than low degree nodes. 
Both linear and superlinear scaling have been observed in real-world networks \cite{Alain}.
The emergence of the weight-topology correlations can be described in the framework of growing network models, including growth of the network by the continuous addition of new links and at the same time an increase on the weights of the existing links driven by a reinforcement dynamics \cite{Weighted}.

Here we study weight-topology correlations in growing simplicial complexes showing that they emerge not just at the network level, but also for $\delta$-faces of higher dimension.
Finally we compare the results obtained in the mean-field approximation with extensive numerical simulations.

The paper is organized as follows. In Sec. II we define weighted simplicial complexes and their main structural properties. In Sec. III we define our model of growing weighted simplicial complexes. In Sec. IV we discuss the mean-field solution of the model. In Sec. V we compare our theoretical prediction  with the results of the numerical simulations. Finally in Sec VI we give the conclusions.

    
\section{Simplicial complexes}
Let us consider $N$ nodes $i=1,2,\ldots, N$.
A simplex of dimension $d$ represents an interaction between a set of $d+1$ of these nodes. 
For instance a  simplex $\alpha$ of dimension $d$ (also called $d$-simplex ) is given by  \bea
\alpha=[i_0,i_1,\ldots, i_d],
\eea
where $i_n$ with $n\in\{0,1,\ldots, d\}$ indicates a node of the simplex.
A  face $\alpha'$ of the $d$-simplex  $\alpha$ is a $\delta$-simplex with $0\leq\delta<d$ formed by a subset of the  nodes of $\alpha$, i.e. $\alpha'\subset \alpha$.
A simplex also has a geometrical interpretation and can be considered as a $d$-dimensional volume. This justifies the choice of calling its subsets  its `faces' .
For example, a simplex of dimension $d=2$ is a triangle, and  all its links and nodes form its faces. Similarly a  simplex of dimension $d=3$ is a tetrahedron and its  faces include four triangles, six links and six nodes. 

A simplicial complex ${\mathcal K}$ of dimension $d$ is a collection of simplices of at most dimension $d$ glued along their shared faces. Additionally every simplicial complex $\mathcal K$ satisfies the following constraint: if a simplex belongs to the simplicial complex (i.e. $\alpha \in {\mathcal K}$) then the simplicial complex also includes all of the faces $\alpha'\subset \alpha$ of that simplex (i.e. $\alpha'\in {\mathcal K}$). In order words the simplicial complex is closed under the operation of inclusion of faces of its simplices.

 In this paper we indicate with $ Q_{d,\delta} (N) $ the set of all possible $\delta$ dimensional faces ($\delta$-faces) in a $d$-dimensional simplicial complex formed by $N$ nodes. Additionally we indicate with ${ S}_{d,\delta}$ the set of $\delta$-faces belonging to the simplicial complex under consideration.
 Here we consider exclusively $d$-dimensional simplicial complexes constructed by gluing $d$-dimensional simplices.
 The structure of such $d$-dimensional simplicial complexes of $N$ nodes is  determined by the adjacency tensor $\bf{a}$ with elements $a_{\alpha}=1,0$ indicating whether the simplex  $\alpha \in Q_{d,d} (N)$ is present ($a_{\alpha}=1$) or absent ($a_{\alpha}=0$) from the simplicial complex, i.e.
\bea
a_{\alpha}=\left\{\begin{array}{ccc} 1 & {\mbox{if}}& \alpha\in { S}_{d,d} ,\nonumber \\
0&&\mbox{otherwise.} \end{array}\right.
\eea 
The weights of the simplices are indicated  by the weight tensor $\bf{w}$, with elements $w_\alpha $ indicating the weight of simplex $\alpha$. 
In a simplicial complex representing co-authorship, for example, a simplex represents a set of co-authors that have collaborated on at least one paper together, while the weight of that simplex corresponds to the total number of papers that have been co-authored by the team.
To characterize the properties of the simplicial complex, we use here the
 {\it{generalized degrees}} and {\it{generalized strengths}} of the $\delta$-faces. 

 The generalized degree $k_{d,\delta}^{\alpha}$ of a $\delta$-face $\alpha \in S_{d,\delta}$  \cite{CQNM,NGF,Owen1} is the number of $d$-dimensional simplices incident to it
\bea 
k_{d,\delta}^{\alpha}= \sum_{\alpha' \in Q_{d,d}(N) | \alpha'\supseteq\alpha} a_{\alpha'}.
\eea
The generalized strength $s_{d,\delta}(\alpha) (t)$ of a $\delta$-face $\alpha\in S_{d,\delta} $  is the sum of the weights  of the $d$-dimensional simplices incident to it
\bea 
s_{d,\delta}^{\alpha}= \sum_{\alpha' \in Q_{d,d}(N) | \alpha'\supseteq\alpha} a_{\alpha'} w_{\alpha'}.
\label{sdef}
\eea

The generalized degree of a $\delta-$face $\alpha$ is related to the generalized degree of the $\delta'$-dimensional faces incident to it, with $\delta'>\delta$, by the simple combinatorial relation  \cite{Owen1}
\bea
k_{d,\delta}^\alpha=\frac{1}{\left(\begin{array}{c}d-\delta\\ \delta'-\delta\end{array}\right)}\sum_{\alpha'\in S_{d,\delta'}|\alpha'\supseteq \alpha}k_{d,\delta'}^{\alpha'}.
\label{rel1}
\eea
Moreover, since every $d$-dimensional simplex belongs to $\left(\begin{array}{c}d+1\\ \delta+1\end{array}\right)$ $\delta$-dimensional faces, in a simplicial complex with $M$ $d$-dimensional simplices we have
\bea
\sum_{\alpha\in {\cal S}_{d,\delta}}k_{d,\delta}^\alpha=\left(\begin{array}{c}d+1\\ \delta+1\end{array}\right)M.
\eea
Interestingly the generalized strength a $\delta$-face $\alpha$ satisfies also 
\bea
s_{d,\delta}^\alpha=\frac{1}{\left(\begin{array}{c}d-\delta\\ \delta'-\delta\end{array}\right)}\sum_{\alpha'\in S_{d,\delta'}|\alpha'\supseteq \alpha}s_{d,\delta'}^{\alpha'}.
\label{sgeng}
\eea
The skeleton of a simplicial complex is the network formed by all its $0$-faces (nodes) and $1$-faces (links).
To a weighted simplicial complex we can associated a skeleton that is a weighted network in which the weights $\omega_{ij}$ of the links in the skeleton are equal to the generalized strengths of the links in the simplicial complex i.e. 
\bea \omega_{ij}=s_{d,1}^{[i,j]},\label{omega}\eea and the strength $S_i$ of a node $i$ is 
\bea
S_i=\sum_{j=1}^N \omega_{ij}.
\eea
The generalized degrees   strength $s_{d,0}^{[i]}$ of the node $i$ of the simplicial complex is naturally related with  the strength $S_i$ of the node in the skeleton network. 
Using Eq. $(\ref{sgeng})$ and Eq. $(\ref{omega})$ it is possible to see that 
\bea
S_i=ds_{d,0}^{[i]}.
\eea
Instead in general the only relation between the generalized degree of node $k_{d,0}$ and the  degree $K_i$ of the same node in the skeleton network is  
\bea
K_i\leq d k_{d,0}^{[i]}
\eea
because  some of the $d$-simplices incident to the node $i$ might share some links ($1$-faces).

In weighted networks, it has been shown that it is possible to characterize the interplay between the network topology and the weights of the links by classifying networks depending on the scaling of the strength as a function of the degree of the nodes.
Specifically it has been shown that for some networks the weights of the links are distributed rather uniformly, resulting in a linear dependence of the strength of the nodes with its degree,
\bea
S_i\propto K_i
\eea while in other networks hub nodes tend to have links with higher weights than low degree nodes. This latter scenario results in a superlinear scaling of the strength versus the degree, i.e. 
\bea
S_i\propto (K_i)^{\theta},
\eea
with $\theta>1$.
An example of networks with linear dependence of the strength versus degree are collaboration networks while an example of non-linear dependence of strength on degree are for instance airport networks where the weights measure the number of passengers for each flight connection.
In Ref. \cite{Weighted} it has been shown that a simple growing network model with reinforcement of the links is actually able to generate networks with linear and superlinear scaling of the strength versus degree depending on the rate at which new links are added with respect to the rate at which links are reinforced.

Here we are proposing a model for growing simplicial complexes showing a very rich phenomenology, and we show evidence that in simplicial complexes it is possible to characterize the correlations between weights and topology by exploring the dependence of the generalized strength $s_{d,\delta}^{\alpha}$ versus the generalized degree $k_{d,\delta}^{\alpha}$.
Specifically we are able to predict three alternative possible scalings: linear, superlinear and exponential, i.e. 
\bea
s_{d,\delta}^\alpha \propto \left\{\begin{array}{l} k_{d,\delta}^\alpha , 
\\ \big(k_{d,\delta}^\alpha\big)^\theta , \\
\exp\left[\beta k_{d,\delta}^{\alpha}\right] ,
\end{array}\right.
\eea
with  $\theta>1$ and $\beta$ indicating a constant greater than zero. 
In this case the superlinear scaling indicates weight-topology correlations, and these correlations are even more pronounced for the exponential scaling.
\begin{figure}[ht!]
	\centering
	\includegraphics[width=0.6\columnwidth]{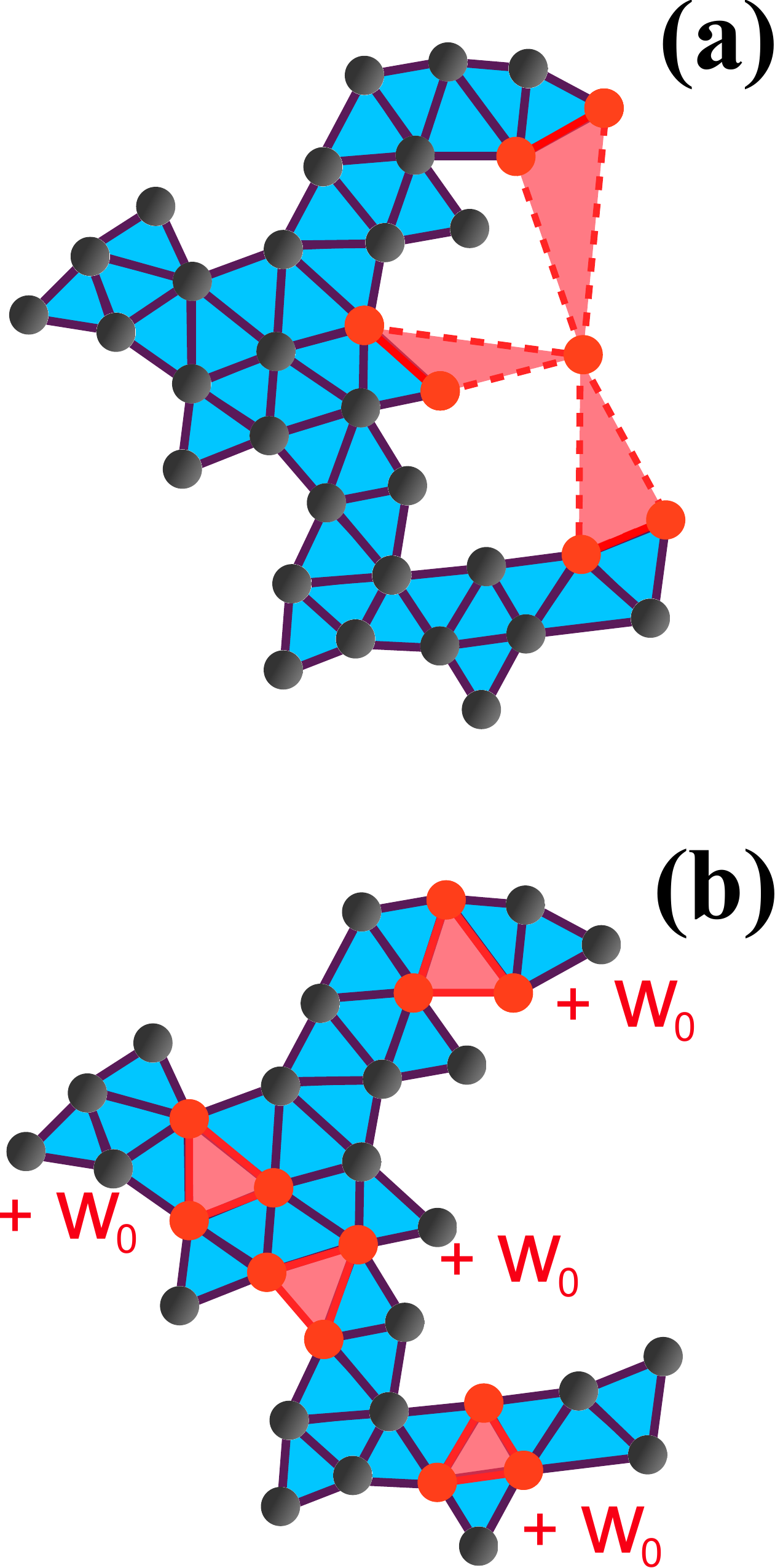}
 	\caption{(Color online) Graphical representation of process A (panel (a)) and process B (panel (b)) for a $2$-dimensional simplicial complex with $m=3$ and $m'=4$ starting from a given initial condition.}
	\label{fig:model}
\end{figure}
\section{The Model} 
\label{model}

In this section we present a model of growing weighted simplicial complexes based on reinforcement of the weights of the simplices and capable of displaying important weight-topology correlations depending on the value of its parameters.
This model is based on the already proposed model of Network Geometry with Flavor \cite{NGF} but includes two important new elements with respect to the mentioned model: i)  the simplicial complexes generated by this model are weighted, ii) the simplicial complexes generated by this model can have non-trivial homology.

In  this model the weighted growing  simplicial complexes are generated as follows. 
We start at time $t=1$ from an initial finite simplicial complex that comprises $m_0>m$ $d$-dimensional simplices of total weight $\omega_{0}$. At each time-step $t>1$ two processes take place:
\begin{itemize}
\item[A)] {\it Add $m$ simplices}: \\A new node arrives and $m$ new $d$-simplices with initial weight $w_0$ are created between the node and pre-existing $(d-1)$-faces. The probability $\Pi_{d-1}(\alpha)$ that a given $(d-1)$-face $\alpha$  is selected is given by
\bea 
\Pi_{d-1}(\alpha) = \frac{1}{\mathcal{Z}_t}  (1 + s n_\alpha),  
\label{ZA}\eea 
where $n_{\alpha}=k_{d,d-1}^{\alpha}-1$ is called the occupancy number and where $s$ is a parameter called  {\em flavor} which takes  the values $s=-1,0,1$ and controls the simplicial complex topology. Note that in Eq. $(\ref{ZA})$,  $\mathcal{Z}_t$ is a normalization constant given by $\mathcal{Z}_t= \sum_{\mu \in S_{d,d-1}(t) } (1+ s n_\alpha)$.
\item[B)] {\it Reinforce $m'$ simplices}: \\At this step $m'$ existing $d$-simplices are selected and their weights are increased by $w_0$. A $d$-simplex $\alpha$ with weight $w_\alpha$ is selected for reinforcement with probability $\tilde{\Pi}_d({\alpha})$ proportional to its weight, i.e.
\bea 
\tilde{\Pi}_d({\alpha }) = \frac{  w_{\alpha} }{\tilde{\mathcal{Z}}_t},
\label{ZB}\eea 
where $\tilde{\mathcal {Z}}_t=\sum_{\alpha \in S_{d,d}(t) } w_{\alpha}$.
\end{itemize}

 In Figure $\ref{fig:model}$ we  describe the two processes (process A and process B) for a $2$-dimensional simplicial complex starting from a given initial condition. 
The flavor $s$ has an important effect on the topological properties of the simplicial complexes produced. Selection of $s=-1$ imposes the constraint that the generalized degree $k_{d,d-1}(\alpha)$ of a $(d-1)$-face $\alpha$ can only take the values $1$ and $2$, or equivalently imposes that $n_{\alpha}$ can only take values $0$ and $1$, which leads to the simplicial complex produced being a $d$-dimensional manifold. Choosing $s=0$ or $s=1$ removes this constraint, and gives a selection probability $\Pi_{d-1}(\alpha)$ that is uniform on the set of all $(d-1)$-faces for $s=0$ and a form of preferential attachment with $\Pi_{d-1}(\alpha) \propto k_{d,d-1}^\alpha$ for $s=1$.

\begin{figure}
	\centering
	\includegraphics[width=0.6\columnwidth]{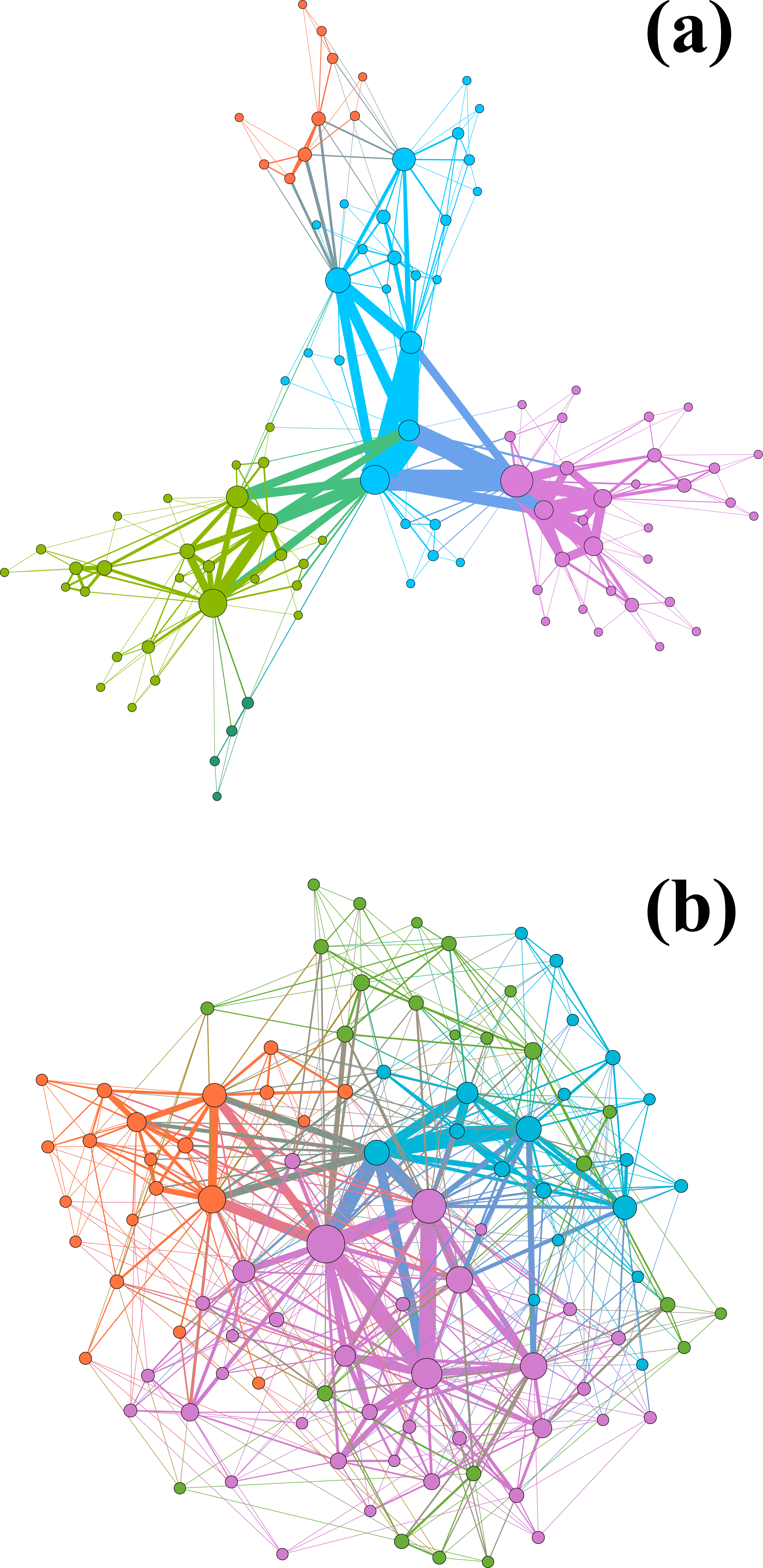}
 	\caption{Skeleton networks of simplicial complexes generated by the model for $d=3$, $N=100$ and $s=-1$. Node sizes indicate their  degrees while link widths indicate their generalized strength. Node and edge colorings indicate community structure calculated according to the Louvain algorithm \cite{Louvain}. Panel (a) shows the skeleton of a simplicial complex with $m=1$ and $m'=2$ while panel (b) shows the skeleton of a simplicial complex with $m=2$ and $m'=1$.}
	\label{fig:gephi}
\end{figure}

In Figure $\ref{fig:gephi}$ we plot the weighted skeleton networks of two simplicial complexes generated by the model in the case $d=3$ and $s=-1$ for $(m,m')=(1,2)$ and $(m,m')=(2,1)$. The weights of the links in these networks indicate the generalized strengths of the links in their corresponding simplicial complexes. While in the case $(m,m')=(1,2)$ nodes with high degree have typically links with stronger weights than the weights of low-degree nodes, the weights are more homogeneously distributed in the case $(m,m')=(2,1)$.
 
We note that for the case  $m=1$ each addition of a new node and its initial $d$-dimensional simplex does not change the Euler characteristics $\chi$, i.e. indicating with $\chi(t)$ the Euler characteristics at time $t$ we get
\bea
\Delta \chi(t)=\chi(t)-\chi(t-1)=0.
\eea
Therefore if the initial condition has $\chi(0)=1$ we obtain $\chi(t)=1$ for every time $t$ and we have a trivial topology.
 However for the case $m>1$ the Euler characteristics change with time according to 
 \bea
 \Delta \chi(t)=1-m<0,
 \eea
 and therefore it could be very interesting to study in more detail the topology of these networks. Additionally the information on the weights could be exploited, by considering, as in Ref. \cite{Vaccarino1,Vaccarino2}, the persistent homology induced by a weights-based filtration  of the simplicial complex.

\section{Mean-field solution of the model}
\subsection{Mean-field solutions for the generalized degrees} \label{GDMF}

In this section our aim is to derive the time evolution of the generalized degrees of the $\delta$-faces of the simplicial complexes using a mean-field approximation.
To this end, let us define  the probability 
$\Pi_{\delta}(\alpha)$  that due to the addition of a single new simplex in the simplicial complex (process A) the $\delta$-face $\alpha$ increases its generalized degree.
The probability $\Pi_{\delta} (\alpha) $ is the sum of the probabilities that any $(d-1)$ face $\alpha'\supseteq \alpha$ is chosen for attaching a new simplex, i.e.

\bea
\Pi_{\delta} (\alpha) &=&\sum_{\alpha'\in S_{d,d-1}|\alpha'\supseteq \alpha}\Pi_{d-1}(\alpha')\nonumber \\
&=&\frac{1}{{\mathcal Z}_t}\sum_{\alpha'\in S_{d,d-1}|\alpha'\supseteq \alpha}1-s+sk_{d,d-1}^{\alpha'} ,
\eea

To derive this expression explicitly in terms of the generalized degree of $\alpha$ we use the following relation
 \bea
 \sum_{\alpha'\in S_{d,d-1}|\alpha'\supseteq \alpha}1-s+sk_{d,d-1}^{\alpha'}=
 (1-s)c_{\delta} \nonumber \\
 + \big(d + s - \delta -1\big)k_{d,\delta}^{\alpha},
 \label{Pia0}
 \eea
where 
\bea
c_{\delta}=\left\{\begin{array}{lll}1 &\mbox{for}& \delta>0 ,\\
m&\mbox{for}& \delta=0 .\end{array}\right.
\eea 
To see why Eq. $(\ref{Pia0})$ holds, observe that Eq. $(\ref{rel1})$ implies the following
\bea  
\sum_{\alpha' \in S_{d,d-1} | \alpha'\supset\alpha}k_{d,d-1}^{\alpha'} = \big(d-\delta\big)k_{d,\delta}^{\alpha}. \eea 
Additionally the following relation can also be shown to hold for our model
\bea  
\sum_{\alpha' \in S_{d,d-1} | \alpha'\supset\alpha} 1 =\left\{\begin{array}{lll}1 + \big(d-\delta-1\big)k_{d,\delta}^{\alpha} &\mbox{for}&{\delta>0},\\
m + \big(d-1\big)k_{d,\delta}^{\alpha}&\mbox{for}& \delta=0.  \end{array}\right.\label{uno}\eea 
This equation can be derived by observing that each initial $d$-simplex which includes the $\delta$-face $\alpha$ contributes by $d-\delta$ to the sum on the left hand side of Eq. $(\ref{uno})$. In fact there are $\left(\begin{array}{c} d-\delta\\d-\delta-1\end{array}\right)=d-\delta$ ways to choose the  $d-(1+\delta)$ nodes of a $(d-1)$-face that do not belong to the $\delta$-face out of the $d-\delta$ nodes of the $d$-simplex that do not belong to the $\delta$-face. Initially any $\delta$-face with $\delta>0$ belongs to a single simplex, while the $\delta$-faces with $\delta=0$ (the nodes) belong to $m$ simplices.
Finally  every $d$-simplex that further increases the generalized degree of the $\delta$-face contributes to the sum just by $d-\delta-1$ because simplices are glued along $(d-1)$-faces of the simplicial complex.

By using  Eq. $(\ref{Pia0})$, we can express $\Pi_{\delta}(\alpha)$ in terms of the generalized degree $k_{d,\delta}^{\alpha}$, as
\bea  
\Pi_{\delta} (\alpha) = 
\frac{(1-s) c_{\delta}+ \big(d + s - \delta -1\big)k^{\alpha}_{d,\delta}}{{\mathcal Z}_t}. 
\label{Pia} 
 \eea 

From this expression the mean-field equations for the generalized degree can be easily derived.
In the mean-field approximation the generalized degrees, of the $\delta$-faces are approximated with their average value over different stochastics realization of the simplicial complex. Additionally, using a very well established framework for simple networks \cite{Doro_book,NS,Newman_book}, we will consider a continuous time approximation in which the (average) degree $k_{d,\delta}(t,t_{\alpha})$ that a $\delta$-face $\alpha$ arrived in the network at time $t_{\alpha}$ has at time $t$ is determined by deterministic differential equations.
These equations read for any $0\leq \delta\leq d-1$,
\bea
   \frac{\partial}{\partial t}  k_{d,\delta}(t,t_{\alpha}) =  m\Pi_{\delta} (\alpha)   ,
\label{MFgd}
\eea
where $\Pi_{\delta}(\alpha)$ is given by Eq. $(\ref{Pia})$.
Let us now note that the  normalization constant ${\mathcal Z}_t$ is simply given, in the limit $t\gg1$ by 
\bea
{\mathcal Z}_t&=&\sum_{\alpha'\in S_{d,d-1}}1-s+sk_{d,d-1}^{\alpha'}\simeq m(d+s)t.
\label{Za}
\eea
In fact the total number of $(d-1)$-faces $\sum_{\alpha'\in S_{d,d-1}}1\simeq mdt$ for $t\gg1$ because at each time we add $m$ new $d$-dimensional simplices each one  contributing  $d$ new $(d-1)$-faces. Additionally
 we have that $\sum_{\alpha'\in S_{d,d-1}}k_{d,d-1}^{\alpha'}\simeq m(d+1)t$ for $t\gg1 $ because any new simplex increases by one the generalized degree of each of its $(d+1)$ $(d-1)$-faces.
Therefore using Eq. $(\ref{Za})$ and Eq. $(\ref{Pia})$ we can derive that, for sufficiently large times, the mean-field equation determining the generalized degree dynamics is given by, 
\bea  
\frac{\partial k_{d,\delta}(t,t_{\alpha})}{d t}=\frac{(1-s)c_{\delta} +\big(d+s-\delta-1\big)k_{d,\delta}(t,t_{\alpha})}{\big(d+s\big)t}, 
\label{MFgd2}
 \eea  
 with initial condition 
 \bea
 k_{d,\delta}(t_{\alpha},t_{\alpha})=c_{\delta}=\left\{\begin{array}{lll}1 &\mbox{for}& \delta>0 , \\m&\mbox{for} &\delta=0 . \end{array}\right.
 \eea
The solution of this equation is 
\begin{widetext}
\bea  
k_{d,\delta}(t,t_{\alpha})= \left\{\begin{array}{lll}
	c_{\delta}\frac{d-\delta}{d+s -\delta -1}\bigg(\frac{t}{t_\alpha}\bigg)^{\lambda_{\delta}} + c_{\delta}\frac{s-1}{d+s -\delta -1} &\mbox{for } &\delta- s\neq d-1, \\
	c_{\delta}\frac{1-s}{d+s}\log\bigg(\frac{t}{t_\alpha}\bigg) + c_{\delta} &\mbox{for} &\delta- s = d-1,
\end{array}\right.
\label{gendegreeMF}
 \eea  
 \end{widetext}
 with 
 \bea
 \lambda_{\delta}=\frac{d+s-\delta-1}{d+s}.
 \label{lambda_delta}
 \eea

The generalized degree distribution $P_{d,\delta}(k)$, for $\delta=d-1$ and $s=-1$ is bimodal, because only the generalized degrees $k_{d,d-1}^{\alpha}=1,2$ are allowed. For all the other cases it is possible to derive the generalized degree distribution using the mean-field solution given by Eq. $(\ref{gendegreeMF})$.
In this way it  is found that the generalized degree distribution is   exponential for $d-1+s-\delta=0$ and power-law for $d-1+s-\delta>0$.
In order to derive these results let us note that since at each time we add a constant number of $\delta$-faces,  the explicit expression for the probability $\hat{P}_{\delta}(t_{\alpha}<\tau)$ that a random $\delta$-face has been  added at time $t_{\alpha}<\tau$, is given by 
\bea
\hat{P}_{\delta}(t_{\alpha}<\tau)=\frac{\tau}{t}.
\eea
Using this result and Eq. $(\ref{gendegreeMF})$   the probability that the generalized degree $k_{d,\delta}(t,t_{\alpha})$ is greater than $k$ can be calculated to be  given by 
\begin{widetext}
\bea
P(k_{d,\delta}(t,t_{\alpha})\geq k)=\left\{\begin{array}{lll}
\Big(  \frac{c_\delta(d-\delta)}{k(d+s-\delta-1)} \Big)^{\frac{1}{\lambda_{\delta}}} &\mbox{for } &\delta- s< d-1,
\\ \exp\Big[-\frac{d+s}{(1-s)c_{\delta}}k \Big] &\mbox{for} &\delta- s = d-1.
\end{array}\right. 
\eea 
\end{widetext}
This leads to the following generalized degree distribution $P_{d,\delta}(k)$ 
\begin{widetext}
\bea
P_{d,\delta}(k)=-\frac{dP(k_{d,\delta}(t,t_{\alpha})\geq k)}{dk}=\left\{\begin{array}{lll}
\frac{d+s}{d+s-\delta-1}\Big(  c_\delta \frac{d-\delta}{d+s-\delta-1} \Big)^{\frac{1}{\lambda_{\delta}}} k^{-\frac{1}{\lambda_{\delta}} -1} &\mbox{for } &\delta- s< d-1,
\\ \frac{d+s}{(1-s)c_{\delta}}\exp\Big[-\frac{d+s}{(1-s)c_{\delta}}k \Big] &\mbox{for} &\delta- s = d-1,
\end{array}\right. 
\label{Pddelta}
\eea
\end{widetext}
valid as long as  $\delta-s< d$.
Therefore the generalized degree distribution of $\delta$-dimensional simplices in growing simplicial networks with flavor $s$ follows Table $\ref{table1}$.
Additionally the  generalized degree distribution $P_{d,\delta}(k)$ given by Eq. $(\ref{Pddelta})$  decays as a power-law $P_{d,\delta}(k)\propto k^{-\gamma_{d,\delta}}$ with a power-law exponent  
\bea
\gamma_{d,\delta}=1+\frac{1}{\lambda_{\delta}}=1+\frac{d+s}{d+s-\delta-1}
\eea
as long as  $\delta-s<d-1$. These distributions are scale-free if $\gamma_{d,\delta}\leq 3$, or equivalently they are scale-free if 
\bea
d\geq d_c^{[\delta,s]}=2(\delta+1)-s.
\eea
Let us now observe  in the considered growing simplicial complex the degree of a node $K_i$  not belonging to the set of nodes in the initial condition, is given by 
\bea
K_i=k_{d,0}^{i}+(d-1)m.
\eea
In fact  initially each node has degree $dm$, and subsequently the degree increases by one for any $d$-simplex glued  to one of the $(d-1)$-faces of the node.
It follows than that if the generalized degree distribution of the nodes is scale-free then the degree distribution of the skeleton network is also scale-free. As a result,  growing simplicial complexes of flavor $s=1$ are  scale-free for any $d\geq1$, the ones of flavor $s=0$ are scale-free for $d\geq2$ and the ones of flavor $s=-1$ are scale-free for $d\geq 3$.
\begin{table}
\center
\label{table1}
\caption{\label{table1} Distribution of generalized degrees of faces of dimension $\delta$ in a $d$-dimensional NGF of flavor $s$ at $\beta=0$.  For $d\geq d_c^{[\delta,s]}=2(\delta+1)-s$ the power-law distributions are scale-free, i.e.  the second moment of the distribution diverges. }
\footnotesize
\begin{tabular}{@{}llll}
\hline
flavor &$s=-1$&$s=0$&$s=1$\\
\hline
$\delta=d-1$&Bimodal&Exponential&Power-law\\
\hline
$\delta=d-2$&Exponential&Power-law& Power-law\\
\hline
$\delta\leq d-3$&Power-law&Power-law& Power-law\\
\hline
\end{tabular}\\
\end{table}
\begin{figure*}
	\centering
	\includegraphics[width=1.8\columnwidth]{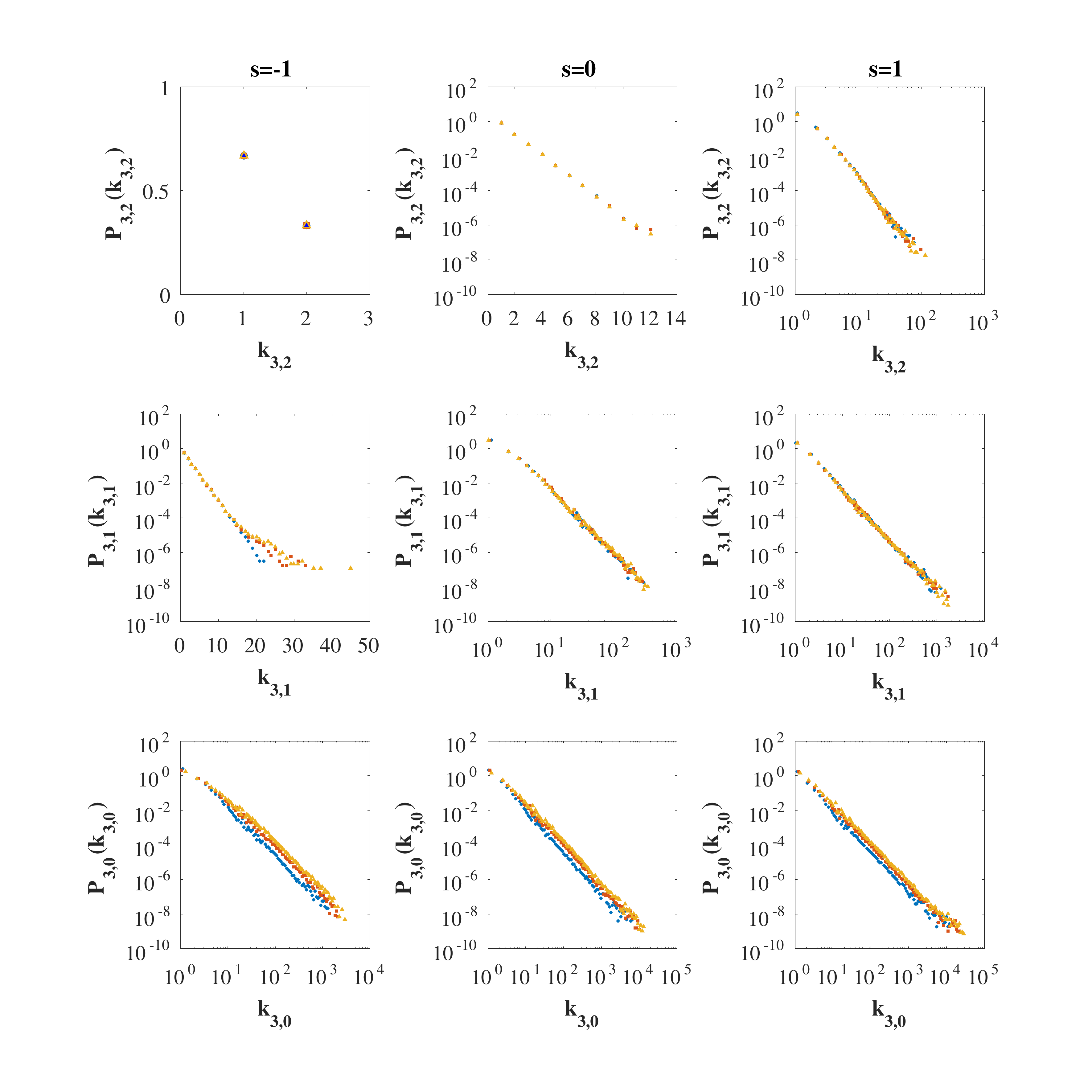}
 	\caption{Generalized degree distributions $P_{d,\delta}(k_{d,\delta})$
 	are shown for simplicial complexes of dimension $d=3$, and  flavor $s=-1$ (left column), $s=0$ (center column) and $s=1$ (right column) for different  $\delta$-faces.  The results of simulations are shown  for $m=1,2$ and $3$ (blue circles, red squares and yellow triangles respectively). The simulated simplicial complexes have $N=10^5$ nodes and the results are averaged over 10 simplicial complexes realizations.}
	\label{fig:pk}
\end{figure*}

\subsection{Probability of a simplex} \label{simp_prob}
In this section we derive the probability of a $\delta$-simplex in terms of the arrival times of its nodes.  

Let us indicate   each $\delta$-face $\alpha_{\delta}$ as the sequence of its nodes $\alpha_{\delta}=[j_0 , j_1 , ... , j_{\delta} ]$ where the nodes are ordered according to the time of their arrival in the simplicial complex, i.e. $t_{j_0} < t_{j_1} < ... < t_{j_{\delta}}$. 
A new node appears in the simplicial complex at every time-step and forms $d$-dimensional simplices with $m$ already existing $(d-1)$-faces. Thus, the $\delta$-face $\alpha_{\delta}$ is the result of the subsequent addition of new $d$-dimensional simplices to the faces $\alpha_{\delta'}=[j_0 , j_1 , ... , j_{\delta'} ]$ formed by the $\delta'+1$ oldest nodes of $\alpha_{\delta}$ for each $0\leq \delta' < \delta$.
 Specifically, after node $j_0$ is added to the network at time $t_{j_0}$ we must have that the node $j_1$ which arrived at the simplicial complex at time $t_{j_1}$ belongs to a new $d$-simplex  incident to the  node $j_0$. Subsequently node $j_2$ which arrived in the simplicial complex at time $t_{j_2}$ should belong to a $d$-simplex incident to the face $\{j_0,j_1\}$ and so on.
Therefore  probability $p_{\alpha_{\delta}}$  that the $\delta$-face $\alpha_\delta$ belongs to the simplicial complex may be written
\bea  \label{deltaprob1}  p_{\alpha_\delta} =\prod_{n=0}^{\delta-1} \pi_{n}(t_{j_{n+1}},t_{j_{n}}),
\eea 
where $\pi_{n}(t_{j_{n+1}},t_{j_{n}}) $ is the probability that the $j_{n+1}$ node arrived in the simplicial complex at time $t_{j_{n+1}}$ belongs to a $d$-simplex incident to the face $\alpha_{n}$ formed by the set of nodes $\{j_0,j_1,\ldots, j_n\}$ of arrival times $t_{j_0} < t_{j_1} < ... < t_{j_{n}}$.
 
Let us observe that in  the mean-field approximation, as we have shown in the previous section,  the generalized degree $k_{d,\delta}^{\alpha_{\delta}}$ of the $\delta$-face $\alpha_{\delta}$ only depends on the time $t_{j_{\delta}}$ of arrival  of the younger node of the simplex, i.e. $k_{d,\delta}^{\alpha_{\delta}}=k_{d,\delta}(t,t_{j_\delta})$. Therefore $\pi_{\delta}(t_{j_{\delta+1}},t_{j_\delta})$ is given by  
\bea 
\pi_{\delta}(t_{j_{\delta+1}},t_{j_\delta})  
&  =&\frac{(1-s)c_{\delta} + \big(d + s - \delta -1\big)k_{d,\delta}(t_{j_{\delta+1}},t_{j_\delta}) }{\big(d+s\big)t_{j_{\delta+1}}},  \nonumber 
\label{pis}
\eea 
where we have used $\pi_{\delta}(t_{j_{\delta+1}},t_{j_\delta})=m\Pi_{\delta}(\alpha_{\delta})$ and the expression of $\Pi_{\delta}$ given by Eqs. (\ref{Pia}) and (\ref{Za}).
Replacing $k^{\alpha_\delta}_{d,\delta}(t_{j_{\delta+1}})$  with the mean-field (expected) generalized degree $k_{d,\delta}(t_{j_{\delta+1}},t_{j_\delta}) $ given by Eq. $(\ref{gendegreeMF})$ we obtain 
\bea  \label{pmu1}
\pi_{\delta}(t_{j_{\delta+1}},t_{j_\delta}) =  c_{\delta}\frac{d-\delta}{d+s}{t_{j_\delta}}^{\frac{ 1+\delta}{d + s} - 1}t_{j_{\delta+1}}^{-\frac{1+\delta}{d + s} }.
\eea

Finally, using Eq. (\ref{pmu1}) and Eq. $(\ref{deltaprob1})$ we get a closed expression for the probability $p_{\alpha_{\delta}}$ of a $\delta$-face as a function of the times $\{t_{j_1}, t_{j_2},\ldots, t_{j_{\delta}}\}$ of arrival of its nodes in the simplicial complex, given by 
\bea  p_{\alpha_\delta}  = m \frac{d!}{(d-\delta)! (d+s)^\delta} \big(t_{j_0} t_{j_1} ... t_{j_{\delta - 1}} \big)^{\frac{1}{d+s} -1} t^{-\frac{\delta}{d+s}}_{j_\delta}  .
\label{ptdep}
\eea 
\subsection{Mean-field solution for the weight of a simplex} \label{WMF}

Here we derive a mean-field expression for the (average) weight $w(t,t_{\alpha})$ that the $d$- dimensional simplex  $\alpha$ added to the simplicial complex at  time $t_\alpha$ has at time $t$.\\\\
Since, according to process B at each time we reinforce $m'$ random simplices increasing their weight by $w_0$, we have 
\bea   \frac{\partial w(t,t_{\alpha})}{\partial t} = w_0 m'\tilde{\Pi}_d({\alpha})  ,
\label{weightDE}
 \eea 
where  $\tilde{\Pi}_d({\alpha })=w_{\alpha}(t)/{\tilde{\mathcal Z}_t}$ is the probability that the $d$-simplex $\alpha$ is reinforced at time $t$.
This equation has initial condition 
\bea
w(t_{\alpha},t_{\alpha})=w_0 
\label{ini_w}
\eea
since each new simplex initially has weight $w_0$.
At each time $m$ new simplices, each of weight $w_0$, are added to the simplicial complex, and $m'$ existing simplices increase their weight by $w_0$.
Therefore we have that the  normalization constant $\tilde{\mathcal {Z}}_t$ is given by 
\bea
 \tilde{\mathcal {Z}}_t=\sum_{\alpha \in S_{d,d}(t) } w_{\alpha} (t)= (m'+m)w_0t+\omega_0\simeq (m'+m)w_0t , \nonumber
 \eea
 where the last expression is valid  for $t\gg 1$.
 It results that the mean-field Eq. $(\ref{weightDE})$ for the weights can be also written as 
\bea
\frac{\partial w(t,t_{\alpha})}{\partial t} =\lambda \frac{w(t,t_{\alpha})}{t} ,
\eea
where 
\bea
\lambda=\frac{m'}{m+m'}.
\label{lambda}
\eea
Given the initial condition expressed by Eq. $(\ref{ini_w})$, this  equation has  solution
\bea  w(t,t_{\alpha}) =  w_0 \left(\frac{t}{t_\alpha}\right)^\lambda.  \label{weightMF} 
 \eea  

\subsection{Mean-field approach for the  generalized strengths} \label{SMF}

In this section we evaluate the generalized strength of a $\delta$-face in the  mean-field approximation. 
In the spirit of the mean-field approximation, i.e. neglecting fluctuations,  we identify the generalized strength $s_{d,\delta}^\alpha $ with its expected value $s_{d,\delta}(t,t_{\alpha})$ over different simplicial complex realizations and conditioned on the existence of the face $\alpha$ with arrival time $t_{\alpha}$. This is given by 
\bea  s_{d,\delta}(t,t_{\alpha}) =  \frac{\sum_{\alpha' \in Q_{d,d}(N) | \alpha'\supset\alpha} p_{\alpha'} w(t,t_{\alpha'})}{p_{\alpha}}. 
\label{gsm}\eea 
Let us indicate each $\delta$-simplex $\alpha$ by the set of its nodes $[i_0, i_1, \ldots , i_{\delta}]$ ordered according to the arrival times in the simplicial complex $t_{i_0}< t_{i_1}< \ldots <t_{i_\delta}=t_{\alpha}$.
Similarly we will indicate each $d$-simplex $\alpha'$ by the ordered set of its nodes $[j_0,j_1, \ldots ,j_d]$ ordered according to the arrival times in the simplicial complex $t_{j_0}< t_{j_1}< \ldots <t_{j_d}=t_{\alpha'}$.
When the $\delta$-simplex $\alpha$ is a face of the $d$-simplex $\alpha'$ we have 
$[i_0,i_1,\ldots i_{\delta}]\subset [j_0,j_1,\ldots j_n]$.
In this case we may indicate with $q(r)$ the index of the node  $i_r\subset \alpha$ in the list $[j_0,j_1,\ldots j_n]$ specifying the nodes of the face $\alpha'$. Therefore we have 
\bea
i_r=j_{q(r)}.
\eea
To be concrete let us consider an example.  In a simplicial complex of dimension $d=4$  consider the 4-simplex $\alpha'$
\bea
\alpha'=[j_0,j_1,j_2,j_3,j_4]=[5,7,11,19,25]
\eea
and the 1-face $\alpha$
\bea
\alpha=[i_0,i_1]=[7,19].
\eea
Since $i_0=j_1$ and $i_1=j_3$ we have 
\bea
q(0)=1, & q(1)=3.
\eea
 In  Eq. $(\ref{gsm})$ let us now distinguish between contributions to the average generalized strength $s_{d,\delta}(t,t_{\alpha})$ from $d$-simplices which contain the nodes of $\alpha$ in the positions specified by distinct $\{q(r)\}_{r=0,1,\ldots \delta}$. Additionally, noting that  $p_{\alpha'}$ in the mean-field approximation depends only on the set of arrival times of its nodes, and that each node is uniquely identified by its arrival time,  and also taking the continuous approximation for arrival times $t_{j_n}$, we get the following expression for the average  generalized strengths 
\begin{widetext} 
\bea  \label{strength1} s_{d,\delta} (t,t_{\alpha}) =\frac{1}{p_{[i_0,...,i_{\delta}]}}\sum_{\{q(r)\}_{r=0,1,\ldots \delta}}\int_{\substack{t_{j_0} < ... < t_{j_d}}} \left[\prod_{n=0}^d dt_{j_n}\right]\prod_{r=0}^{\delta} \hat{\delta}\left( t_{j_{q(r)}},t_{i_r}\right)p_{[j_0,...,j_d]} w (t,t_{j_d}),  \eea
\end{widetext}
where $\hat{\delta}(x,y)$ indicates the Kronecker delta. 
Using Eq. $(\ref{ptdep})$ for the probability $p_{[j_0,...,j_d]}$ and Eq. $(\ref{weightMF})$ for the analytical expression of $w(t,t_d)$ we get
\bea  \label{strength5} s_{d,\delta} (t,t_{\alpha}) = \frac{1}{p_{\alpha}}w_0 m  \frac{d!}{ (d+s)^d}t^\lambda \big(t_{i_0} t_{i_1} ... t_{i_{\delta}} \big)^{\frac{1}{d+s}-1} \nonumber
\\ \times \sum_{\{q \}}A_{q(\delta)} \left(\prod_{r=0}^{\delta-1} X_{q(r),q(r+1)} \right) B_{q(0)},
\eea
where $A_{q(\delta)}$ is the integral over all arrival times greater than $t_{j_q(\delta)}$, $X_{q(r),q(r+1)} $ is the integral over arrival times between $t_{j_{q(r)}}$ and $t_{j_{q(r+1)}}$, and $B_{q(0)} $ is the integral over arrival times less than $t_{j_{q(0)}}$. 

All of these quantities can be expressed in term of the function $I(\tau,t)$ given by 
\bea
I_{\tau , t}^n &=& \int^{t}_{\tau} dt_n t_{n}^{\frac{1}{d+s} -1}\int^{t_{n}}_{\tau} dt_{n-1} t_{n-1}^{\frac{1}{d+s} -1} ...\int^{t_{2}}_{\tau} dt_{1} t_{1}^{\frac{1}{d+s} -1}.\nonumber
\label{Itaut}
\eea

In particular, by  distinguishing between the cases in which there is at least one node whose arrival time is being integrated over, and the case  where the allocation of positions specified by $\{q\}$ implies  that there are no arrival times to integrate over we obtain

\bea \label{Adelta0} A_{q(\delta)} =  \left\{ \begin{array}{lrc} \int^{t}_{t_{i_\delta}}dt_{j_d} t_{j_d}^{-\lambda -\frac{d}{d+s}} I_{t_{i_\delta},t_{j_d}}^{d- q(\delta) -1} & \text{if} & 0 \leq q(\delta) \leq d-1,
\\ t_{i_\delta}^{-\lambda + \frac{s-1}{d+s}} & \text{if} & q(\delta)=d,
\end{array} \right.  \eea

\bea X_{q(r),q(r+1)}= \left\{ \begin{array}{lrc} I_{t_{i_r},t_{i_{r+1}}}^{q(r+1)- q(r) -1} & \text{if} & q(r+1)-q(r)>1,
\\1 & \text{if} & q(r+1)-q(r)=1,
\end{array} \right. \eea

\bea  B_{q(0)} =  \left\{ \begin{array}{lrc} I_{0,t_{i_0}}^{q(0)} & \text{if} & q(0)>0, \vspace*{3mm}
\\ 1 & \text{if} & q(0)=0.
\end{array} \right.   \eea 

We note here that  Eq. (\ref{strength5}) may be simplified further, by substituting the expression for $p_\alpha$ given in Eq. (\ref{ptdep}):

\bea
s_{d,\delta} (t,t_{\alpha}) =w_0 \frac{(d-\delta)!}{(d+s)^{d-\delta}} t^\lambda t_{i_\delta}^{-\frac{d+s-\delta-1}{d+s}} \nonumber
\\ \times \sum_{\{q \}}A_{q(\delta)} \left(\prod_{r=0}^{\delta-1} X_{q(r),q(r+1)} \right) B_{q(0)}.  \label{strength_sim}\eea 
\begin{figure*}
	\centering
	\includegraphics[width=1.8\columnwidth]{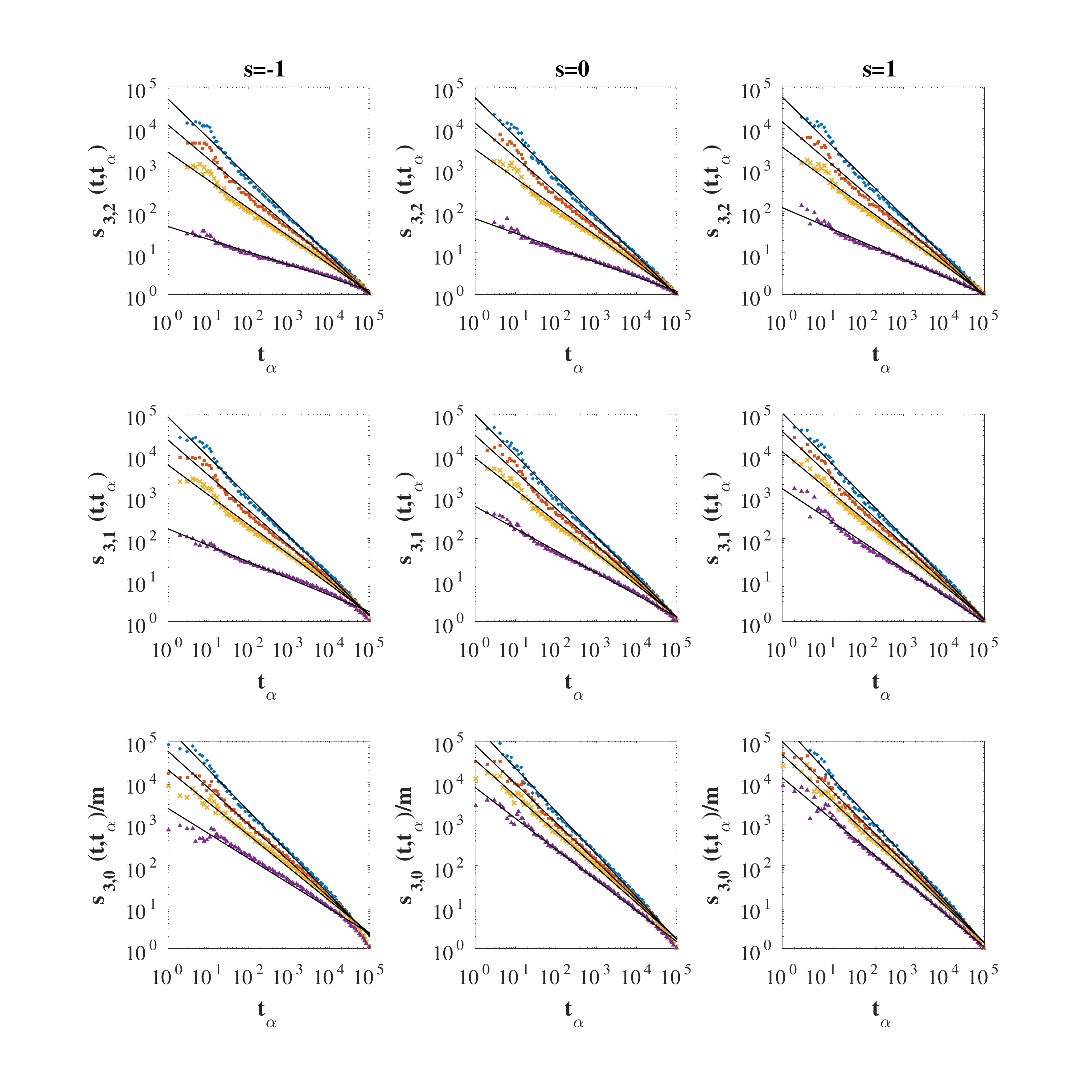}
		\caption{The average generalized strengths  $s_{d,\delta}(t,t_{\alpha})$ of the $\delta$-faces arrived in the network at time $t_{\alpha}$ 
 	are shown for simplicial complexes of dimension $d=3$, and  flavor $s=-1$ (left column), $s=0$ (center column) and $s=1$ (right column).  The results of simulations are shown   for $(m=1$, $m'=5)$, $(m=2$, $m'=5)$, $(m=2$, $m'=3)$ and $(m=3$, $m'=1)$ (blue circles, red squares, yellow x's and purple triangles respectively). The simulated simplicial complexes have $N=10^5$ nodes and the results are averaged over 10 simplicial complexes realizations.}
	\label{fig:st}
\end{figure*}
This expression can be shown  to depend only on the ratio between the time $t$ and the time $t_{\alpha}=t_{i_\delta}$. Specifically it can be shown  (see the Supplementary Material for details of the derivation) that $s_{d,\delta}(t,t_{\alpha})$ is given by 

\begin{widetext}
\bea
s_{d,\delta} (t,t_{\alpha}) = \left\{ \begin{array}{lrc} w_0 \frac{d-\delta}{(d+s)(\lambda_\delta - \lambda)} \Big(\frac{t}{t_{\alpha}}\Big)^{\lambda_\delta} + w_0 \Bigg[1 -  \frac{d-\delta}{(d+s)(\lambda_\delta - \lambda)}\Bigg] \Big(\frac{t}{t_{\alpha}}\Big)^{\lambda} & \text{if} & \lambda\neq\lambda_\delta,
\\w_0 \Big(\frac{t}{t_{\alpha}}\Big)^{\lambda}  \Bigg[ 1 +\frac{d-\delta}{d+s}\log\Big(\frac{t}{t_{\alpha}}\Big) \Bigg]& \text{if} & \lambda = \lambda_\delta,
\end{array}\right.
\label{s} \eea
where $\lambda_{\delta}$ is given by Eq. $(\ref{lambda_delta})$ and $\lambda$ is given by Eq. $(\ref{lambda})$.
\end{widetext}
For $t/t_{\alpha}\gg 1$ keeping only the leading terms  of the above expression we get
\bea
s_{d,\delta} (t,t_{\alpha})\propto  \left\{ \begin{array}{lrc} \Big(\frac{t}{t_{\alpha}}\Big)^{\lambda_\delta} & \text{if} & \lambda<\lambda_\delta,\nonumber \\
\Big(\frac{t}{t_{\alpha}}\Big)^{\lambda} & \text{if} & \lambda>\lambda_\delta,
\\
 \Big(\frac{t}{t_{\alpha}}\Big)^{\lambda}  \log\Big(\frac{t}{t_{\alpha}}\Big) & \text{if} & \lambda = \lambda_\delta.
\end{array}\right.
\label{slead} \eea

\begin{figure}
	\centering
	\includegraphics[width=0.6\columnwidth]{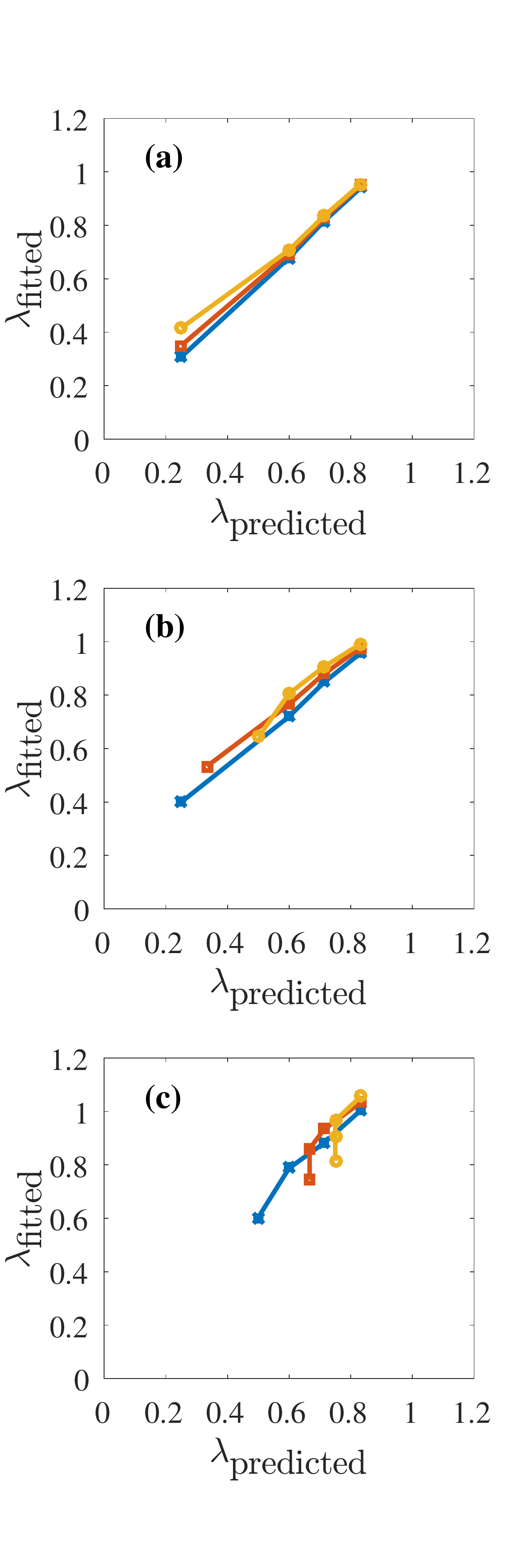}
 	\caption{The exponents  $\lambda_{fitted}$ obtained by fitting Eq. (\ref{fittingeq}) to the data in figure \ref{fig:st} are shown versus the predicted exponents $\lambda_{predicted}$ given by Eq. \ref{fitted} for different $\delta$-faces. The panels (a), (b) and (c) refer respectively to triangles ($\delta=2$),  links $(\delta=1)$ and nodes ($\delta=0$). The blue stars, red squares, and yellow circles indicate the data obtained respectively for the the flavors  $s=-1$, $s=0$ and $s=1$.}
	\label{fig:compare}
\end{figure}
Finally we can evaluate the scaling of the average generalized strength versus the generalized degree $s_{d,\delta}(k_{d,\delta})$ using the mean-field approximation.
To this end we  keep only the leading terms for $t/t_{\alpha}\gg 1$ both in Eq. $(\ref{gendegreeMF})$ for the average generalized degrees $k_{d,\delta}(t,t_{\alpha})$ and in Eq. $(\ref{s})$ for the average generalized strengths $s_{d,\delta}(t,t_{\alpha})$ and we neglect the fluctuations of the generalized degrees ($k_{d,\delta}(t,t_{\alpha})\simeq k_{d,\delta}^\alpha$) and generalized strengths ($s_{d,\delta}(t,t_{\alpha})\simeq s_{d,\delta}^\alpha$). As long as  $ \lambda_{\delta}>0$, we obtain
\bea \label{scaling}
{s}_{d,\delta} (k_{d,\delta})\propto\left\{\begin{array}{lrc}  k_{d,\delta}& \mbox{for}& \lambda<\lambda_{\delta}, \\
 k_{d,\delta} \ln  k_{d,\delta} & \mbox{for}& \lambda=\lambda_{\delta}, \\
\left(k_{d,\delta}\right)^{{\lambda}/{\lambda_{\delta}}} & \mbox{for}& \lambda>\lambda_{\delta}.
\end{array}\right.
\eea
For $\lambda_{\delta}=0$, instead we derive an exponential scaling of the average of the generalized strength versus the average of the generalized degree of the $\delta$-faces, i.e.
\bea \label{scaling2}
{s}_{d,\delta}(k_{d,\delta})\propto  e^{\beta k_{d,\delta}},
\eea
with $\beta= \lambda \frac{d+s}{(1-s)c_{\delta}}$. 
These results predict that  by tuning the parameter values $m$ and $m'$  (determining $\lambda$ as for Eq. $(\ref{lambda})$)  is possible to observe either linear, superlinear of even exponential scaling of the generalized strengths versus the generalized degrees.
 
We stress here that the scaling relations Eqs. $(\ref{scaling})$ and $(\ref{scaling2})$ are obtained  in the limit $t/t_{\alpha}\gg 1$ neglecting the  fluctuations of the generalized degrees and the generalized strengths over different network realizations.  Therefore these expressions need to be compared to numerical simulations for assessing the limits of the considered approximations.

\begin{figure*}
	\centering
	\includegraphics[width=1.8\columnwidth]{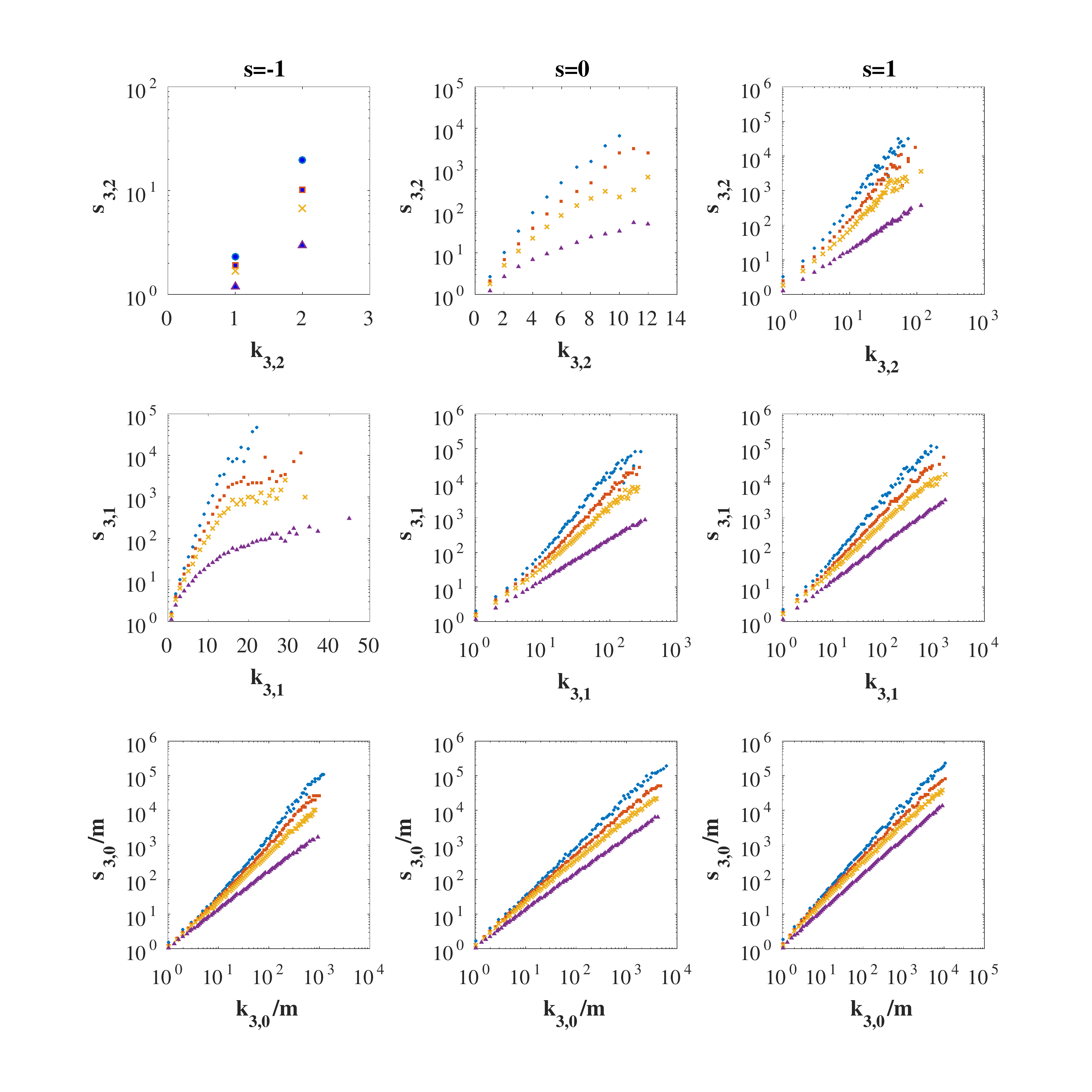}
		\caption{The average generalized strengths  of $\delta$-faces as a function of their corresponding generalized degree $s_{d,\delta}(k_{d,\delta})$  	are shown for simplicial complexes of dimension $d=3$, and  flavor $s=-1$ (left column), $s=0$ (center column) and $s=1$ (right column).  The results of simulations are shown   for $(m=1$, $m'=5)$, $(m=2$, $m'=5)$, $(m=2$, $m'=3)$ and $(m=3$, $m'=1)$ (blue circles, red squares, yellow x's and purple triangles respectively). The simulated simplicial complexes have $N=10^5$ nodes and the results are averaged over 10 simplicial complexes realizations.}
	\label{fig:sk}
\end{figure*}

\section{Numerical simulations}

To check the validity of our mean-field calculations we have run extensive simulations of the model. 
When writing the program to implement the model,  care should be taken in efficiently storing the information about the composition of the simplicial complex.  In fact for a simplical complex of $N$ nodes  the total number of potential  $\delta'$-simplices scales like $N^{\delta'+1}$ therefore maintaining adjacency tensors with entries for every potential $\delta'$-simplex would require storing $N^{\delta'+1}$ integers.  Since in  our model, the actual numbers of $d$-simplices and  $\delta$-faces scales linearly with $N$ we can  handle efficiently the information about the simplicial complex structure  by keeping a list of the simplices and the faces generated by the model rather than maintaining adjacency tensors.

Here we report and discuss in particular the simulation results obtained for simplicial complexes of dimension $d=3$ with all the possible values of the flavor $s=-1,0,1$ and a variety of choices of $m$ and $m'$.

Our main goal is to characterize the limit of validity of the mean-field calculations performed in the previous section.

In Figure $\ref{fig:pk}$ we report the simulation results for the generalized degree distribution $P_{d,\delta}(k_{d,\delta})$ and $N=10^5$ averaged over 10 realizations of the model.
We observe that the mean-field calculation predicts exactly for which dimension $\delta$ and for which flavor $s$ we observe binomial, exponential or power-law distribution. 

In Figure $\ref{fig:st}$ we display the average generalized strengths $s_{d,\delta}(t,t_{\alpha})$ of $\delta$-faces $\alpha$ as a function of their arrival time $t_{\alpha}$.
We observe a clear power-law scaling of $s_{d,\delta}(t,t_{\alpha})$ as a function of $t/t_{\alpha}$ for $t/t_{\alpha}\gg 1$ as predicted by the mean field approximation (Eq. $\ref{slead}$).
In order to evaluate more in detail the limits of validity of the mean-field equations we performed the power-law fits 
\bea
s_{d,\delta}(t,t_{\alpha})=a \left(\frac{t}{t_{\alpha}}\right)^{\lambda_{fitted}},
\label{fittingeq}\eea
valid for $t/t_{\alpha}\gg 1$,
and we have compared the fitted exponent $\lambda_{fitted}$ with the predicted exponent obtained from Eq. (\ref{slead})
\bea
\lambda_{predicted}\simeq\max(\lambda,\lambda_{\delta}).
\label{fitted}\eea 
The comparison between the exponents $\lambda_{fitted}$ and $\lambda_{predicted}$ is shown in Figure $\ref{fig:compare}$ for simplicial complexes of dimension $d=3$ and different values of $m$ and $m'$ determining $\lambda$.
We observe that while the overall trend of $\lambda_{fitted}$ is captured by the mean-field result, some deviations are observed. These deviations become more significant for $\lambda \simeq \lambda_{\delta}$ where it is expected to be more difficult to observe the leading term in Eq. $(\ref{s})$ starting for finite time simulations results.

Finally, as  discussed in section \ref{SMF}, for simplicial complexes with a large number of nodes the mean-field approximation predicts that   the generalized strengths of the faces are related to their generalized degrees by the scaling relation given in Eqs. (\ref{scaling}) and (\ref{scaling2}). Nevertheless we have discussed that this approximation neglects the role of fluctuations for the generalized degree and for the generalized strengths. Therefore it is important to check to what extend the mean-field calculations capture the simulation results. 
In Figure $\ref{fig:sk}$ we show the average generalized strength versus the generalized degree $s_{d,\delta}(k_{d,\delta})$.
We observe that the role of fluctuations is particularly pronounced for $\delta$-faces with exponential generalized degree distributions (i.e. $\delta=d-2$ for $s=-1$ and $\delta=d-1$ for $s=0$). These fluctuations are more significant for values of the parameters $m$ and $m'$ corresponding to low-values of $\lambda$.
Instead we observe that for the other $\delta$-faces the mean-field predictions provides a rather good  prediction of the scaling of the average generalized strength $s_{d,\delta}(k_{d,\delta})$. \\
 
\section{Conclusions}

In this paper we have presented a non-equilibrium model for weighted  simplicial complexes. In this model simplicial complexes  evolve at each time A) by the addition of a new node belonging to $m$ new $d$-dimensional simplices and B)  by the reinforcement of the weights of $m'$ $d$-dimensional simplices.

The model generates simplicial complexes with non-trivial topology including manifolds,  with either constant Euler characteristics (for $m=1$) or with continuously decreasing Euler characteristics.
The skeleton of these simplicial complexes is a network with notable complex structure, including high clustering coefficient (ensured by the simplicial complex structure) and heterogeneous scale-free degree distribution for $d\geq 3$.

Here we have focused on the rich interplay between topological properties of the simplicial complexes and the distribution of the weights of the simplices.

We have found that is possible to extend the  strength versus degree analysis performed in simple networks to simplicial complexes by characterizing the functional relation between the generalized strengths of their faces versus the corresponding generalized degrees.

Specifically the proposed model is able to generate simplicial complexes where the generalized strength grows linearly, superlinearly or exponentially as a function of the parameters $m$ and $m'$ of the model.

We believe that this model could be rather fruitful for modelling real-world simplicial complexes such as collaboration networks that are typically weighted.
Additionally the model could be used a benchmark to test the wide range of topological and geometrical measures and computational techniques that have been proposed in recent years for the study of real datasets. These include different definitions of curvature, and the persistent homology conducted using a filtration based on the weights of the links or on the simplices.

\clearpage

\onecolumngrid
\clearpage
\renewcommand\theequation{{S-\arabic{equation}}}
\renewcommand\thetable{{S-\Roman{table}}}
\renewcommand\thefigure{{S-\arabic{figure}}}
\setcounter{equation}{0}
\setcounter{figure}{0}
\setcounter{section}{0}

\section*{\Large SUPPLEMENTARY MATERIAL}

\subsection*{Introduction}
In this Supplementary Material we give details of the derivation of Eq.(55) of the main body of the paper expressing the average generalized strength $s_{d,\delta}(t,t_{\alpha})$ of a $\delta$-face $\alpha$ arrived in the network at time $t_{\alpha}$ as a function of time $t$.
\subsection*{Main steps of the derivation of the Eq.(55) }
\label{appendix A}
In the main body of the paper we have derived the following equation (Eq. (54)) for the average generalized strengths $s_{d,\delta}(t,t_{\alpha})$ of the $\delta$-face $\alpha$ with arrival time $t_{\alpha}$
\bea
s_{d,\delta} (t,t_{\alpha}) =w_0 \frac{(d-\delta)!}{(d+s)^{d-\delta}} t^\lambda t_{i_\delta}^{-\frac{d+s-\delta-1}{d+s}}  \sum_{\{q \}}A_{q(\delta)} \left(\prod_{r=0}^{\delta-1} X_{q(r),q(r+1)} \right) B_{q(0)},  \label{strength_sim2}\eea 

where $A_{q(\delta)}$, $B_{q(\delta)}$ and  $X_{q(r),q(r+1)}$ are defined respectively by Eqs. (51), (52) and (53) of the main text.

In order to obtain explicit expressions for $A_{q({\delta}}, X_{q(r),q(r+1)}$ and $B_{q(0)}$ let us observe that  the function  $I_{\tau , t}^n$ defined as

\bea
I_{\tau , t}^n &=& \int^{t}_{\tau} dt_n t_{n}^{\frac{1}{d+s} -1}\int^{t_{n}}_{\tau} dt_{n-1} t_{n-1}^{\frac{1}{d+s} -1} ...\int^{t_{2}}_{\tau} dt_{1} t_{1}^{\frac{1}{d+s} -1}
\label{Itaut2}
\eea
 can be also written as

\bea \label{In0}
I_{\tau, t}^n &=& \frac{(d+s)^n}{n!}\Big(t^{\frac{1}{d+s}} - \tau^{\frac{1}{d+s}} \Big)^n\nonumber \\
&=&\frac{(d+s)^n}{n!}\sum_{r=0}^n \left(\begin{array}{c}n\\r\end{array}\right)(-1)^{r}t^{{\frac{n-r}{d+s}} }\tau^{{\frac{r}{d+s}}}.
\eea

This result allows us to express  $X_{q(r),q(r+1)} $  as

\bea \label{X1} 
\hspace{-5mm}X_{q(r),q(r+1)}&= &\left\{ \begin{array}{lrc} I_{t_{i_r},t_{i_{r+1}}}^{q(r+1)- q(r) -1} & \text{if} & q(r+1)-q(r)>1,
\\1 & \text{if} & q(r+1)-q(r)=1,
\end{array} \right. \nonumber \\
&=&\left\{ \begin{array}{lrc}\frac{(d+s)^{q(r+1)-q(r)-1}}{(q(r+1)-q(r)-1)!}\Big(t_{i_{r+1}}^{\frac{1}{d+s}} - t_{i_r}^{\frac{1}{d+s}} \Big)^{q(r+1)-q(r)-1} & \text{if} & q(r+1)-q(r)>1,
\\1 & \text{if} & q(r+1)-q(r)=1.
\end{array} \right.
\eea
Similarly  $B_{q(0)} $ can be expressed as
\bea \label{B1}  B_{q(0)} &=&\left\{ \begin{array}{lrc} I_{0,t_{i_0}}^{q(0)} & \text{if} & q(0)>0 ,\vspace*{3mm}
\\ 1 & \text{if} & q(0)=0,
\end{array} \right.    \nonumber \\
&=&\left\{ \begin{array}{lrc} \frac{(d+s)^{q(0)}}{q(0)!} t_{i_0}^{\frac{q(0)}{d+s}}&\mbox{if} & q(0)>0, \vspace*{3mm}
 \\ 1 & \mbox{if} & q(0)=0 .\end{array}\right.
\eea 

Finally using Eq. (\ref{In0}) and the definition of $A_{q(\delta)}$ that we rewrite here for convenience,
\bea \label{Adelta02} A_{q(\delta)} =  \left\{ \begin{array}{lrc} \int^{t}_{t_{i_\delta}}dt_{j_d} t_{j_d}^{-\lambda -\frac{d}{d+s}} I_{t_{i_\delta},t_{j_d}}^{d- q(\delta) -1} & \text{if} & 0 \leq q(\delta) \leq d-1, 
\\ t_{i_\delta}^{-\lambda + \frac{s-1}{d+s}} & \mbox{if} & q(\delta)=d,
\end{array} \right.  \eea
 we obtain 
\bea \label{Adelta0} A_{q(\delta)} =  \left\{ \begin{array}{lrc} \frac{(d+s)^{d-q(\delta)-1}}{(d-q(\delta)-1)!} \sum_{r=0}^{d-q(\delta)-1}{ {d-q(\delta) -1}\choose{r}}(-1)^r t_{i_\delta}^{\frac{r}{d+s}} \int^{t}_{t_{i_\delta}}dt_{j_d} t_{j_d}^{-\lambda +\frac{d+s -q(\delta) -r -1}{d+s} -1}    & \text{if} & 0 \leq q(\delta) \leq d-1,
\\ t_{i_\delta}^{-\lambda + \frac{s-1}{d+s}} & \text{if} & q(\delta)=d.
\end{array} \right.  \eea

 In the case $q(\delta) \leq d-1$  the integral present in Eq. (\ref{Adelta0}) has two separate expressions  for $ \lambda = \frac{d +s-q(\delta) -1 -r}{d+s}$ and $\lambda \neq \frac{d +s-q(\delta) -1 -r}{d+s} $.
By performing the integral we find the following expression for $A_{q(\delta)}$ 
\bea \label{Aqd2}  A_{q(\delta)} =  \left\{ \begin{array}{lll} (d+s)^{d-q(\delta)-1} \sum_{r=0}^{d-q(\delta)-1} {D}_{q(\delta) + r} t_{i_\delta}^{\frac{r}{d+s}} \frac{(-1)^r}{r!}  & \text{if} & 0 \leq q(\delta) \leq d-1, 
\\ t_{i_\delta}^{-\lambda + \frac{s-1}{d+s}} & \text{if} & q(\delta)=d,
\end{array} \right.  \eea

where the quantities $D_{q(\delta)+r}$ are given by 

\bea \label{Ahat1} D_{q(\delta) + r}= \left\{ \begin{array}{lrc} \frac{1}{(d-q(\delta) -r -1)!}\Big(\frac{d +s -q(\delta) -1 -r}{d + s} - \lambda \Big)^{-1}  \big(t^{\frac{d +s-q(\delta) -1 -r}{d+s} - \lambda} - t_{i_\delta}^{\frac{d +s-q(\delta) -1 -r}{d+s} - \lambda} \big) & \text{ if } & \lambda \neq \frac{d +s-q(\delta) -1 -r}{d+s},
\\ \frac{1}{(d-q(\delta) -r -1)!} \log \Big( \frac{t}{t_{i_\delta}} \Big) & \text{ if } & \lambda = \frac{d +s-q(\delta) -1 -r}{d+s},
\end{array}\right.\eea
and depend on $q(\delta)$ and $r$ only through their sum. 

In order to calculate the  generalized strengths given by Eq.(\ref{strength_sim2}), let us observe that

\bea \sum_{\{ q \} }A_{q(\delta)} \left(\prod_{r=0}^{\delta-1} X_{q(r),q(r+1)}\right)  B_{q(0)}=\sum_{q(\delta)=\delta}^d A_{q(\delta)}R_{q(\delta),\delta},
\label{Ruse}
\eea
where $R_{q(r),r}$ are functions defined recursively by the following pair of equations, 
\bea 
R_{q(1),1}&=&\sum_{q(0)=0}^{q(1)-1} X_{q(0),q(1)}  B_{q(0)}, \label{aR1}\\
R_{q(\beta),\beta}&=&\sum_{q(\beta-1)=\beta-1}^{q(\beta)-1} X_{q(\beta-1),q(\beta)}R_{q(\beta-1),\beta-1}.\label{bR1}
\eea
The solution of these equations (see next section for details of this derivation) reads
\bea \label{Rhfinal} R_{q(\beta),\beta} = (d+s)^{q(\beta)-\beta}\frac{t_{i_\beta}^{\frac{q(\beta)-\beta}{d+s}}}{(q(\beta)-\beta)!}.
\eea
Therefore  the average generalized strength $s_{d,\delta}(t,t_{\alpha})$ may be written as
\bea s_{d,\delta} (t,t_{\alpha}) =  w_0 \frac{(d-\delta)!}{(d+s)^{d-\delta}} t^\lambda t_{i_\delta}^{-\frac{d+s-\delta-1}{d+s}}\left[A_d R_{d,\delta} + \sum_{q(\delta)=\delta}^{d-1} A_{q(\delta)} R_{q(\delta)} \right].
\label{sR}
\eea

Using  Eq. (\ref{Ahat1}) and Eq. (\ref{Rhfinal})  we get
\bea \label{bracket}  
\sum_{q(\delta)=\delta}^{d-1} A_{q(\delta)}R_{q(\delta),\delta}&=&(d+s)^{d-\delta-1}\sum_{q(\delta)=\delta}^{d-1}\sum_{r=0}^{d-q(\delta)-1} D_{q(\delta) + r} t_{i_\delta}^{\frac{q(\delta) +r -\delta}{d+s}}  \frac{(-1)^r}{r! (q(\delta)-\delta)!}	\nonumber\\
&=& (d+s)^{d-\delta-1} D_{q(\delta) + r} t_{i_\delta}^{\frac{q(\delta) +r -\delta}{d+s}} \Bigg|_{q(\delta) + r = \delta} = (d+s)^{d-\delta-1} {D}_{\delta}.
\eea
Note that in deriving the Eq.(\ref{bracket}) we have used the following mathematical identity
\bea \label{ideq} \sum_{x=a}^{b} \sum_{y=0}^{b - x} f(x+y) \frac{(-1)^{y}}{y! (x-a)!}= f(a),
\eea
valid for some integers $a,b>0$ with $a<b$.
Therefore  the average generalized strength given by Eq. $(\ref{sR})$ can be written as
\bea s_{d,\delta} (t,t_{\alpha})& =&  w_0 \frac{(d-\delta)!}{(d+s)^{d-\delta}} t^\lambda t_{i_\delta}^{-\frac{d+s-\delta-1}{d+s}}\left[A_d R_{d,\delta} + \sum_{q(\delta)=\delta}^{d-1} A_{q(\delta)} R_{q(\delta)} \right]
\\ &=& w_0 \frac{(d-\delta)!}{(d+s)^{d-\delta}} t^\lambda t_{i_\delta}^{-\frac{d+s-\delta-1}{d+s}}\left[\frac{(d+s)^{d-\delta}}{(d-\delta)!}t_{i_\delta}^{-\lambda + \frac{d+s-\delta-1}{d+s}}  + (d+s)^{d-\delta-1} {D}_{\delta} \right],
\eea

which simplifies to

\bea \label{SwithD}
s_{d,\delta}(t,t_{\alpha})= w_0 \Big(\frac{t}{t_{i_\delta}}\Big)^{\lambda} + w_0 \frac{(d-\delta)!}{d+s} t^\lambda t_{i_\delta}^{-\frac{d+s-\delta-1}{d+s}} D_\delta.
\eea

As noted earlier, $D_\delta$ takes different forms in the cases $\lambda \neq \lambda_\delta=\frac{d+s-\delta-1}{d+s}$ and $\lambda = \lambda_\delta=\frac{d+s-\delta-1}{d+s}$. Inserting Eq. (\ref{Ahat1}) into Eq. (\ref{SwithD}) leads to our final expression for the generalized strength:
\bea
s_{d,\delta}^{\alpha} (t) = \left\{ \begin{array}{lrc} w_0 \frac{d-\delta}{(d+s)(\lambda_\delta - \lambda)} \Big(\frac{t}{t_{i_\delta}}\Big)^{\lambda_\delta} + w_0 \Bigg[1 -  \frac{d-\delta}{(d+s)(\lambda_\delta - \lambda)}\Bigg] \Big(\frac{t}{t_{i_\delta}}\Big)^{\lambda} & \text{if} & \lambda\neq\lambda_\delta,
\\w_0 \Big(\frac{t}{t_{i_\delta}}\Big)^{\lambda}  \Bigg[ 1 +\frac{d-\delta}{d+s}\log\Big(\frac{t}{t_{i_\delta}}\Big) \Bigg]& \text{if} & \lambda = \lambda_\delta.
\end{array}\right.
\eea
Since this equation is the same as Eq. (54) of the main text, this concludes here our discussion.

\subsection*{Derivation of Eq. (\ref{Rhfinal})}
\label{AppendixB}

In this section our goal is to show that Eq. (\ref{Rhfinal}) holds.
This equation is given by 
\bea \label{Rhfinal2} R_{q(\beta),\beta} = (d+s)^{q(\beta)-\beta}\frac{t_{i_\beta}^{\frac{q(\beta)-\beta}{d+s}}}{(q(\beta)-\beta)!},
\eea
where $R_{q(r),r}$ are functions defined recursively by the following pair of equations
\bea 
R_{q(1),1}&=&\sum_{q(0)=0}^{q(1)-1} X_{q(0),q(1)}  B_{q(0)},\label{R1b} \\
R_{q(\beta),\beta}&=&\sum_{q(\beta-1)=\beta-1}^{q(\beta)-1} X_{q(\beta-1),q(\beta)}R_{q(\beta-1),\beta-1}.\label{R1c}
\eea

To this end  we first check that (\ref{Rhfinal2}) holds for $\beta=1$. Inserting Eq. (\ref{X1}) for $X_{q(0),q(1)} $ and Eq. (\ref{B1}) for $B_{q(0)}$ into Eq. (\ref{R1b}) gives

\bea R_{q(1),1}= \sum_{q(0)=0}^{q(1)-1} \sum_{l_{0} =0}^{q(1)-q(0)-1} \frac{(d+s)^{q(1) -1} (-1)^{l_0}}{(q(1)-q(0)-1-l_0)!(l_0)!q(0)!}   t_{i_{0}}^{\frac{q(0)+l_0}{d+s}}  t_{i_{1}}^{\frac{q(1)-q(0)-1-l_0}{d+s}}. 
\eea

We note that the expression being summed over factorises into a term depending on $q(0)$ and $l_{0}$ only through their sum and a term depending on $q(0)$ and $l_{0}$ otherwise:

\bea \label{R11} R_{q(1),1} = \sum_{q(0)=0}^{q(1)-1} \sum_{l_{0} =0}^{q(1)-q(0)-1} f(q(0)+l_{0})\frac{(-1)^{l_0}}{l_0!q(0)!},
\eea
where
\bea f(q(0)+l_{0})=(d+s)^{q(1) -1} \frac{t_{i_{0}}^{\frac{q(0)+l_0}{d+s}}  t_{i_{1}}^{\frac{q(1)-q(0)-1-l_0}{d+s}}}{(q(1)-q(0)-1-l_0)!}.
\eea
Using the mathematical identity  Eq.(\ref{ideq}),  Eq. (\ref{R11})  simplifies to
\bea R_{q(1),1}  = f(0)  =(d+s)^{q(1) -1} \frac{ t_{i_{1}}^{\frac{q(1)-1}{d+s}}}{(q(1)-1)!}. 
\eea
So (\ref{Rhfinal2}) holds in the case $\beta=1$. 
\\We now show that in general if Eq. (\ref{Rhfinal2}) holds for some $\beta$ then it must also hold for $\beta+1$. Substituting in Eq. (\ref{X1}) and Eq. (\ref{Rhfinal2}) into Eq. (\ref{R1c}) gives
\bea \label{Rbeta1} R_{q(\beta+1),\beta+1} = \sum_{q(\beta)=\beta}^{q(\beta+1)-1} \sum_{l_{\beta} =0}^{q(\beta+1)-q(\beta)-1}  \frac{(d+s)^{q(\beta+1) - \beta -1}(-1)^{l_{\beta}}}{(q(\beta+1)-q(\beta)-1-l_{\beta})!(l_{\beta})!(q(\beta)-\beta)!}   t_{i_{\beta}}^{\frac{q(\beta)+l_{\beta}-\beta}{d+s}}  t_{i_{\beta+1}}^{\frac{q(\beta+1)-q(\beta)-1-l_{\beta}}{d+s}}
.\eea

Similar to the $\beta=1$ case we may write (\ref{Rbeta1}) in the form
\bea \label{Rhgeneral2} R_{q(\beta+1),\beta+1} = \sum_{q(\beta)=\beta}^{q(\beta+1)-1}  \sum_{l_{\beta} =0}^{q(\beta+1)-q(\beta)-1}  f(q(\beta)+l_{\beta})\frac{(-1)^{l_\beta}}{l_\beta!q(\beta)!},
\eea
where in this case the term depending only on $q(\beta)$ and $l_\beta$ through the sum of the two is
\bea f(q(\beta)+l_{\beta})=(d+s)^{q(\beta+1) - \beta -1} \frac{t_{i_{\beta}}^{\frac{q(\beta)+l_\beta -\beta}{d+s}}  t_{i_{\beta+1}}^{\frac{q(\beta+1)-q(\beta)-1-l_\beta}{d+s}}}{(q(\beta+1)-q(\beta)-1-l_\beta)!}.
\eea
Using the identity (\ref{ideq}) allows us to make the simplification
\bea R_{q(\beta+1),\beta+1} = f(\beta) =(d+s)^{q(\beta+1) - \beta -1} \frac{t_{i_{\beta+1}}^{\frac{q(\beta+1)-\beta-1}{d+s}}}{(q(\beta+1)-\beta-1)!},
\eea
which confirms Eq.(\ref{Rhfinal2}) or equivalently, Eq.(\ref{Rhfinal}).

\end{document}